\documentstyle[12pt,aasms4,psfig]{article}

\newcommand{\ud}[2]{\mbox{$^{+ #1}_{- #2}$}}

\begin{document}

\title{The Proper Motion, Parallax, and Origin of the Isolated Neutron Star
  RX~J185635-3754}
\author{Frederick M.\ Walter}
\affil{Department of Physics and Astronomy, State University of New York, Stony
Brook NY 11794-3800\\fwalter@astro.sunysb.edu}

\begin{abstract}
 
The isolated neutron star RX~J185635-3754 is the closest known neutron star to
the Sun. Based on HST WFPC2 obervations over a 3 year baseline,
I report its proper motion (332$\pm$1 mas~yr$^{-1}$ at a position angle of
100.3$^\circ\pm$0.1$^\circ$) and parallax (16.5$\pm$2.3~mas; 61~pc). This
proper motion brings the neutron star from the general vicinity of the 
Sco-Cen OB association. For an assumed neutron star radial velocity between
$-$55 and $-$60~km~s$^{-1}$,
the runaway O star $\zeta$~Oph, the Upper Sco OB association, and the
neutron star come into spatial coincidence between 0.9 and 1.0 million
years ago.
RX~J185635-3754 may be the remnant of the original primary of the $\zeta$~Oph
system. If so, the space velocity suggests that the neutron star received
a kick of about 200~km~s$^{-1}$ at birth.

\end{abstract}


\section{Introduction}

Most neutron stars known in our galaxy have been identified either
because they are accreting from a binary companion
or because their rapid rotation and strong magnetic field
produces a radio or X/$\gamma$-ray pulsar. 
Studies of accreting neutron stars or pulsars tell us far
more about accretion and magnetospheric physics than they do about the
intrinsic properties of neutron stars.
Based either on pulsar birthrates (Narayan \& Ostriker 1990),
or on the number of supernovae needed to account for the heavy element
abundance of the galaxy (Arnett, Schramm, \& Truran 1989), there are 10$^8$ to
10$^9$ neutron stars in the galaxy. 
There are a few hundred 
accreting neutron star binaries currently known, and about 1000 radio pulsars.
The accreting neutron stars and the
pulsars are not representative of the vast majority of the
neutron stars in the galaxy.

Representatives of the non-accreting, non-pulsing
variety of neutron stars have been eagerly sought
over the past decade (e.g., Greiner 1996; Manning, Jeffries, \& Willmore 1996; 
Belloni, Zampieri, \& Campana 1997),
and some likely candidates have been identified.
These radio quiet neutron stars (see
Caraveo, Bignami, \& Tr\"umper 1996 for a review) form an inhomogeneous
group. 

Some radio quiet neutron stars are
associated with young supernova remnants. These may be radio-quiet because
their radio emission is beamed (e.g., Brazier \& Johnston 1999), they
are highly magnetized yet slowly rotating (e.g., Vasisht et al.\/ 1997), or 
because they are not highly magnetized at birth because the fallback submerges
the field (Geppert, Page, \& Zannias 1999). Examples of young, radio quiet
neutron stars include RX~J0822-4300 in Pup~A (Gaensler, Bock, \& Stappers
2000) and the central point source in Cas~A (Chakrabarty et al.\/ 2000). These
are still young and hot, and are X-ray-bright, with X-ray luminosities of order
10$^{35-36}$~erg~s$^{-1}$.

Other radio quiet neutron star candidates
are not associated with supernova remnants,
and may be old. Isolated neutron stars cool quickly (e.g., Tsuruta 1998),
within about 1 million years (Myr), so 
most neutron stars should be faint (of order
10$^{31}$~erg~s$^{-1}$ or less),
and difficult to find. A handful of old neutron star candidates have been
discovered as X-ray sources in ROSAT images (cf. Caraveo et al.\/
1996). These are soft X-ray sources with approximately
black body spectra, temperatures (kT) of 10 to 100 eV,
and $\frac{f_X}{f_{opt}}$ ratios in excess of 1000. 
Although some may be magnetars (Thompson \& Duncan 1995; Heyl \& Kulkarni 1998),
others may be bona-fide old, isolated, weakly magnetized neutron stars.
Such old isolated neutron stars afford the opportunity to study the
atmospheres of neutron stars (e.g., Pavlov et al.\/ 1996b,
Rajagopal \& Romani 1996).
It is in principle possible to
determine the composition, surface gravity, and angular diameter of an
isolated neutron star from analysis of its spectrum. This can yield both the
mass M and radius R,
and place contraints on the nuclear equation of state (e.g., Lattimer
\& Prakash 1998, 2000).
While it is possible to study the surfaces of pulsars by observing the unpulsed,
thermal component (e.g., Pavlov, Stringfellow, \& Cordova 1996a),
or to determine the mass from the orbital dynamics of the accreting
matter (Miller, Lamb, \& Psaltis 1998; Kaaret, Ford, \& Chen 1997; Zhang,
Strohmayer \& Swank 1997)
or from modelling X-ray bursts (Haberl \& Titarchuk 1995; Strohmayer et al.\/ 
1997), it is easier to do so 
for an isolated non-pulsing neutron star.

The brightest of these isolated neutron star candidates is RX~J185635-3754 
(Walter, Wolk, \& Neuh\"auser 1996). Its optical
counterpart is a blue object with V$\approx$25.7 (Walter \& Matthews 1997).
On the basis of its 57~eV temperature, nearly thermal
spectrum, and projection against the 
$\approx$130~pc-distant R~CrA molecular cloud (Maracco \& Rydgren 1981),
it would have an emitting
area smaller than about 700~km$^2$ if a blackbody. In the presence of an
atmosphere, the emitting area, or radius, must be larger than the blackbody of
the same temperature.
There is no evidence for X-ray pulsations,
and it is radio-quiet. We observe emission from the entire surface of
a neutron star, and not from a heated polar cap.

The important heating processes in an isolated neutron star are heat loss 
from the hot interior (reviewed by Tsuruta 1998) and
accretion from the interstellar medium (e.g., Blaes \& Madau 1993; Treves
et al.\/
2000). In standard cooling models the surface temperature is expected to
plummet below 10$^6$K within 1~Myr, but various surface reheating schemes
have been proposed (e.g., Larson \& Link 1999) to keep the surface hot
for longer times. 
Heating via accretion from the interstellar medium does not decay with
time (although it will vary with the density of the local interstellar
medium), but if the space
velocity exceeds about 15~km~s$^{-1}$, Bondi-Hoyle accretion from the
interstellar gas should be negligable. 

The uncomplicated, isolated, radio quiet neutron star RX~J185635-3754 
affords the opportunity to study the immaculate surface of a neutron star.
To interpret the spectral energy distribution properly, and to determine its
radius, we need its distance.
Here I report on the astrometric analysis of 3 images of 
RX~J185635-3754 taken with the
Hubble Space Telescope WFPC2 detector over the course of 3 years. These images
reveal the parallax and proper motion of the target.

\section{The Data}

We obtained 3 sets of observations with the HST WFPC2 camera
(Table~\ref{tbl-1}). The first epoch images were described by
Walter \& Matthews (1997).
The images were taken near the times of maximum parallactic displacement,
which occurs on 3 October
and 30 March. The images were dithered using the standard 5.5 pixel
diagonal shift. The first and third observations each consist of 4
images; we obtained 8 images during the second observation. The target is
located near the center of the planetary camera (PC).
I did not analyze the images from the other cameras.
I discuss here the images taken through the F606W filter. We also
obtained images through the F178W, F300W, and F450W filters.
The spectrophotometry will be discussed elsewhere.

\placetable{tbl-1}

\section{Analysis Technique}

I analyzed the re-calibrated images delivered through the Multimission Archive
at STScI (MAST).
For each of the three observations I produced a median-filtered, co-added
image for astrometric and photometric analysis. I did this 
by rebinning the data to twice the nominal scale,
shifting the dithered PC images by the nominal offsets, rebinning the images 
back to the nominal scale, and stacking the images. After 
median-filtering the images to remove cosmic rays, I co-added the images.
All the analysis was performed using software written in the IDL language,
incorporating some software from the IDL astronomy users library
(Landsman 1995).

Forty three objects are common to the three PC images, including the target
(Table~\ref{tbl-2}) but excluding the brightest stars.
Four of these objects appear extended, and are likely to be galaxies.
The approximate R-band magnitudes from aperture photometry range from
18.8 to 26.8. 
In the images, stars C, D, F, L, and M of Walter et al.\/ (1996) appear to
be overexposed. Star 20 of Campana et al.\/ (1997)  is clearly a galaxy; their
star 25 is a visual pair with a separation of 0.2~arcsec.

\placetable{tbl-2}

I measured the positions of the objects by fitting two dimensional
Gaussian and Lorentzian line profile, and by fitting 
one-dimensional Gaussians independently in RA and DEC.
The two dimensional fits utilize the MPFIT2DPEAK IDL
procedure\footnote{http://cow.physics.wisc.edu/\~{}craigm/idl/fitting.html}.
The Lorentzian profile is more sharply peaked than the Gaussian profile, and
is somewhat better matched to the HST PSF (as estimated by the $\chi^2$ values
for the fits). The two dimensional fits employ flat backgrounds.
The advantage of the one-dimensional fits is that I can fit a varying
(up to second order) background. This can improve the quality of the fit in
confused regions, near
bright stars, or near diffraction spikes.

Prior to fitting the data I corrected for the
34$^{th}$ row error (Anderson \& King 1999) and for the
geometric distortions in the PC camera (Holtzman et al.\/ 1995).
These detector coordinates are converted into celestial coordinates using the
astrometric information in the FITS header. The absolute positions depend on
the particular guide stars used for each observation.
The positions listed in Table~\ref{tbl-2} are those measured from the 1996
October 6 image. Relative to that image, the coordinates measured in the
subsequent two images are offset by (+3.4,+0.6) and ($-$0.6,+0.3) arcsec in
RA,DEC. 

The astrometric reductions are done in X-Y coordinates, corrected for
detector distortions. I unrolled the
images using the orientation given in the FITS header, so that X,Y correspond
to RA,DEC. Direct comparison of
the images shows no evidence of any residual roll or changes in the nominal
plate scale of 0.0455 arcsec~pixel$^{-1}$.

The position of a star at epoch $k$, relative to the initial position,
is given by
\begin{center}
X$_k -$ X$_0$ = O$_X$($k$) + PM$_X$ dT$_k$ + $\pi_X$(dT$_k$)/d,\\
Y$_k -$ Y$_0$ = O$_Y$($k$) + PM$_Y$ dT$_k$ + $\pi_Y$(dT$_k$)/d\\
\end{center}
where $X_k$, $Y_k$ are the observed position during observation $k$, O is
the systematic offset between the two coordinate systems, dT$_k$ is the
elapsed time since the first observation, PM is the proper motion, $\pi$ is
the magnitude of the parallactic displacement at one pc, and d is the distance.
The position angle of the major axis of the parallactic ellipse is 83$^\circ$.

I determined the systematic offsets O$_X$,O$_Y$ between the
observations (see above) by taking the
weighted mean of the differences in the X and Y positions after rejecting
positions that are
discrepant from the mean at more than 3$\sigma$.
The weighted mean and median offsets 
agree to better than 2~milli-arcsec.
There was no evidence for any significant deviations from the nominal
roll angle.
Removing the offset leaves a coupled set of 2 equations with two
unknowns in each dimension, which I then solve for the parallax
and proper motion. 

\section{Results}
\begin{figure}
\centerline{
\psfig{figure=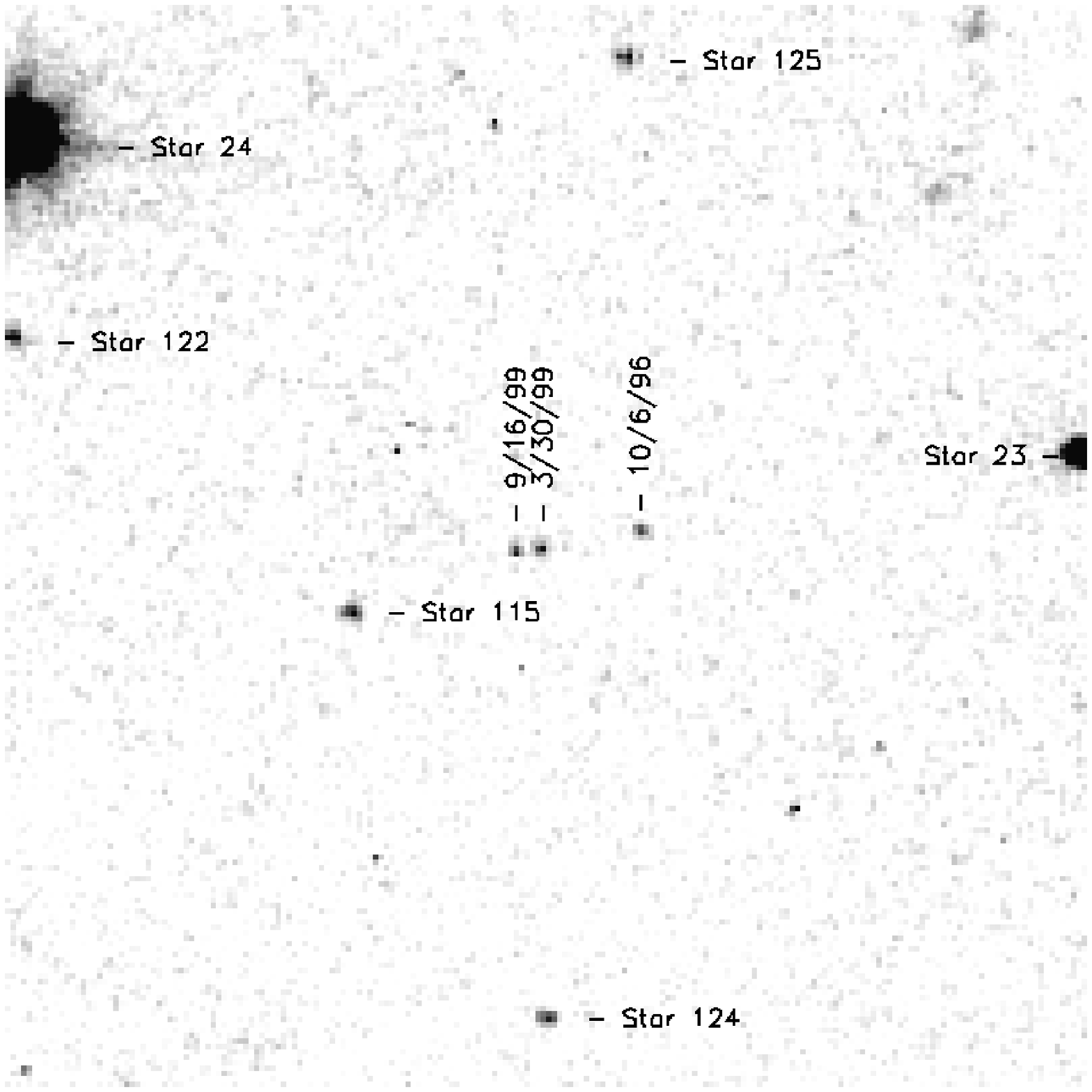,angle=00.,width=8.0in}}
\caption{The motion of RX~J185635-3754. This is an exposure-weighted
sum of the three F606W images.
North is up, east is to the left. The image is 9.1 arcsec
(200 PC pixels) across.
The three images were rotated, coaligned (to the nearest pixel)
by cross-correlation, and co-added. 
The three images of the neutron star are labelled by date.
Other stars in the image are labelled; these are slighly elongated E-W due
to the simple alignment procedure.
}\label{fpmim}
\end{figure}

The measured motions of RX~J185635-3754 are summarized in Table~\ref{tbl-3},
and the motion is shown in Figure~\ref{fpmim}. 
The three sets of measurements yield results which agree within their
uncertainties. In Table~\ref{tbl-3} I quote 68\% and 95\% joint confidence
uncertainties on the derived quantities. To determine the uncertainties,
I ran a Monte Carlo simulation, selecting positions at random from a 
normally-distributed sample
with mean equal to the measured offsets and variance equal to the measurement
error.

\placetable{tbl-3}
\placefigure{fpmim}

\begin{figure}
\centerline{
\psfig{figure=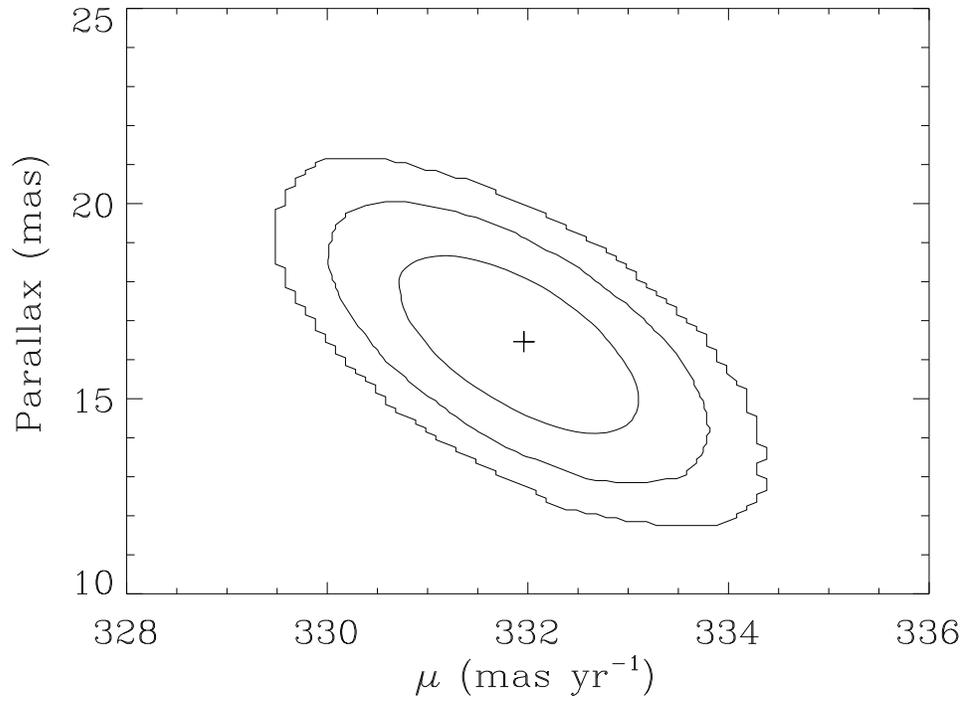,angle=00.,width=6.0in}}
\caption{The joint confidence contours for the proper motion and the
distance. The 68\% (1~$\sigma$), 95\% (2~$\sigma$), and 99.9\% (3~$\sigma$)
confidence contours are plotted; the best estimate is marked with the plus sign.
}\label{fig-cc}
\end{figure}

RX~J185635-3754 star is moving just south of east
at a rate of $\onethird$ of an arcsecond per year\footnote{This proper motion
was first detected using the two ROSAT HRI observations. Walter \& An (1998) 
reported a proper motion of 0.4$\pm$0.2~arcsec~yr$^{-1}$ in a southeasterly
direction.}.
The parallax of 16~milli-arcsec (32 milli-arcsec full amplitude)
is detected at greater than 7$\sigma$.
Note that the directions of the proper motion and the major axis of the
parallactic ellipse are co-aligned to within 17$^\circ$. A 1\% error in the
magnitude of the proper motion yields a 12\% uncertainty in the parallax.
The joint confidence contours (Figure~\ref{fig-cc}) show the correlation.

\placefigure{fig-cc}

\begin{figure}
\centerline{
\psfig{figure=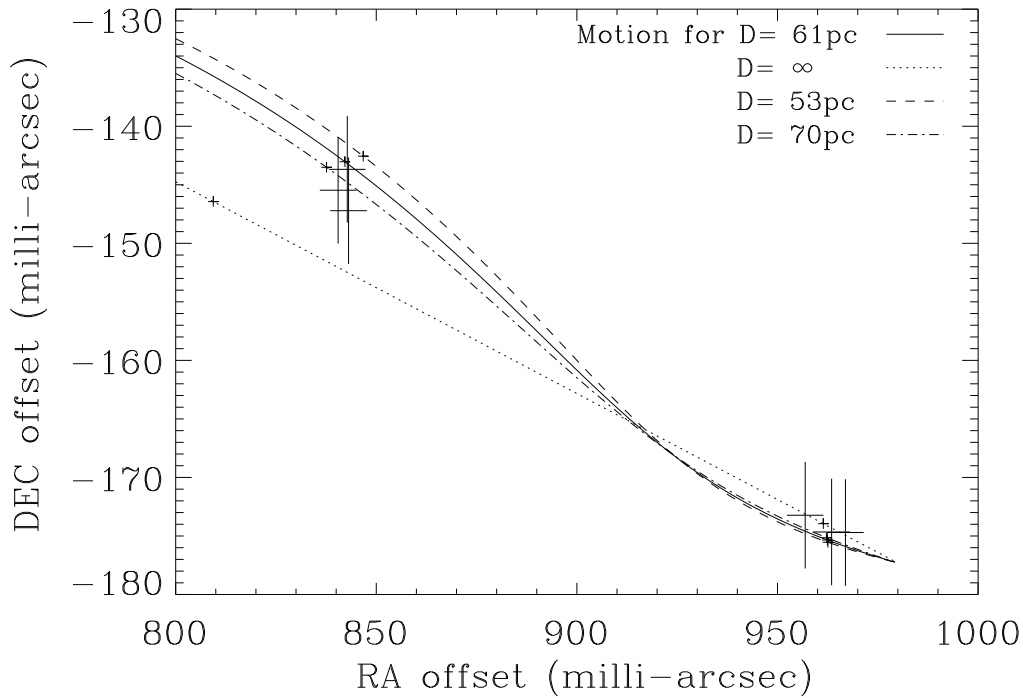,angle=00.,width=6.0in}}
\caption{The best fit proper motion and parallax.
The measured spatial offsets of the target from the initial
observation (0,0) are plotted as error bars. There are 3 measurements for each
observation, as described in the text.
The curves show the predicted position
for the nominal distance of 61~pc, for the $\pm$1~$\sigma$
envelope, and for a source at infinity.
The small crosses show the expected positions of the target on the
day of observation. The motions in RA provide a stronger constraint on the
distance than does the DEC wobble.
Note that the RA and DEC scales are not the same.
}\label{fig-pmpx}
\end{figure}

\begin{figure}
\centerline{
\psfig{figure=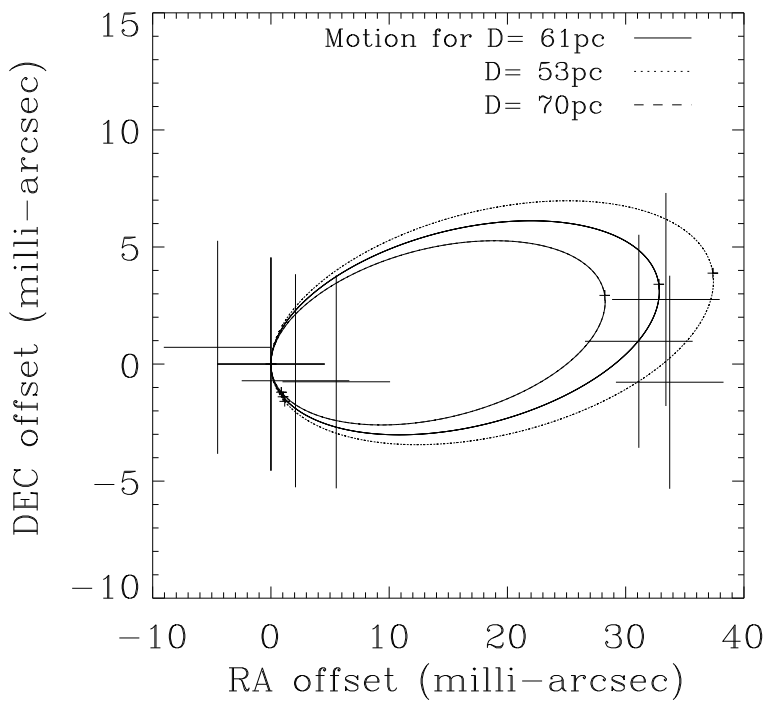,angle=00.,width=6.0in}}
\caption{The parallactic ellipse. The measured
spatial offsets, with the proper motion subtracted, 
are plotted on the predicted parallactic ellipse for
for the nominal distance of 61~pc and for the $\pm$1~$\sigma$
envelope. The large error bars represent the three measurements at each epoch;
the point at (0,0) represents the initial observation.
The small crosses show the expected positions of the target on the
day of observation. Note that the RA axis is twice the length of the DEC axis.
}\label{fig-zpmpx}
\end{figure}

The components of the space motion are illustrated in in Figure~\ref{fig-pmpx}.
Figure~\ref{fig-zpmpx} shows the measured positions and the
parallactic ellipse after subtracting the proper motion.
\placefigure{fig-pmpx}
\placefigure{fig-zpmpx}

Since the determinations of the proper motion and parallax are done with
respect to all the objects in the image, it is possible that they can be biased
by systematic motions of the comparison stars. These would manifest themselves
as systematic offsets in the two mean positions. Given the brightnesses of
the comparison stars, it is unlikely that there will be any detectable 
mean space motions or parallax. The brightest star, with F$_{606W}\approx$18.9,
must be at a distance modulus in excess of 6 if an M dwarf (the $B-R$ color
suggests an F dwarf at a distance modulus in excess of 10, although it could be
a closer white dwarf). The bulk of the
comparison stars, with R$>$23, unless very heavily reddened,
are more distant than 1~kpc if
dwarfs. In this direction (l=0$^\circ$, b=$-$20$^\circ$),
it is likely that we are
seeing mostly evolved stars in the galactic bulge. Consequently it is unlikely
that there is any detectable, systematic motion of the comparison stars.

\section{Discussion}

The proper motion is consistent with the expectation that
RX~J185635-3754 is a nearby neutron star.
The proper motion is sufficiently large that accretion
from the interstellar medium, which scales as v$^{-3}$, produces 
negligable heating of the surface. In the Blaes \& Madau (1993) formalism,
accretion heating for a surface filling factor $f$=1 and an interstellar
density n$_H$=1~cm$^{-3}$ can account for only about 1\% of the observed
luminosity of the star.

\subsection{Parallax and Implications for the Radius}

The distance of 61 \ud{9}{8} pc is small;
this is likely the
nearest known neutron star to the Earth. The distance is an important
parameter to know well,
since both the space velocity and the inferred radius scale with the distance.

Walter et al.\/ (1996) claimed a firm upper
limit to the distance of about 130~pc,
based on the low X-ray absorption column of 1.4$\times$10$^{20}$~cm$^{-2}$
(E$_{B-V}$=0.02~mag).
Even though our analysis of the multi-wavelength spectrum (An 2000; Pons et al.,
in preparation)
suggests a higher column of about 2$\times$10$^{20}$~cm$^{-2}$, this is still
comfortably less than that expected from the CrA molecular cloud in this
direction. However, a column of 2$\times$10$^{20}$~cm$^{-2}$ implies a mean
interstellar density of 1~cm$^{-3}$, which is substantially higher than the
mean in the local ISM. This would be worrysome, were it not for the fact that
Knude \& H$\o$g (1998) show that substantial reddening exists in front
of the CrA cloud, with extinctions E$_{B-V}$ of up to 0.1 for stars within
50~pc.

At this distance, the blackbody radius 
R$_\infty$ ($\equiv\frac{\rm R}{\sqrt{1-\frac{2GM}{Rc^2}}}$)
of the object is 3.3~km using the
57~eV temperature reported by Walter et al.\/ (1996). 
The X-ray and optical data are better fit by model
atmospheres of heavy element composition
(An 2000; Walter et al.\/ 2000; Walter \& An 1998; Pons et al., in preparation),
with kT$\sim$49~eV and an angular
diameter R$_\infty$/d=0.18$\pm$0.05. In this case,
R$_\infty$~=~11.2$\pm$3.4~km. I caution that the inferred radius is
model-dependent, and cannot be considered a firm measurement until 
the surface composition is measured observationally from good-quality
X-ray spectra. Nevertheless, the radius suggests that
we are indeed viewing the entire surface of the neutron star, and not just a 
heated polar cap.

\subsection{History of this Neutron Star}
The proper motion of the neutron star provides clues to its origin. If the
age is small enough one might hope to trace the star back to a plausible
birthplace and place constraints on the possible progenitors and the
evolutionary history of this neutron star. Here
the proper motion directs us back towards the Sco-Cen OB association, 
a source of supernovae during the past few million years (de Geus, de Zeeuw,
\& Lub 1989), and the
apparent birthplace of the runaway O star $\zeta$~Oph (Blaauw 1991, 1993).
To clarify the relation, if any, of RX~J185635-3754, the Sco-Cen
OB association, and $\zeta$~Oph, requires running the space motions
of these objects backwards through time.

\begin{figure}
\centerline{
\psfig{figure=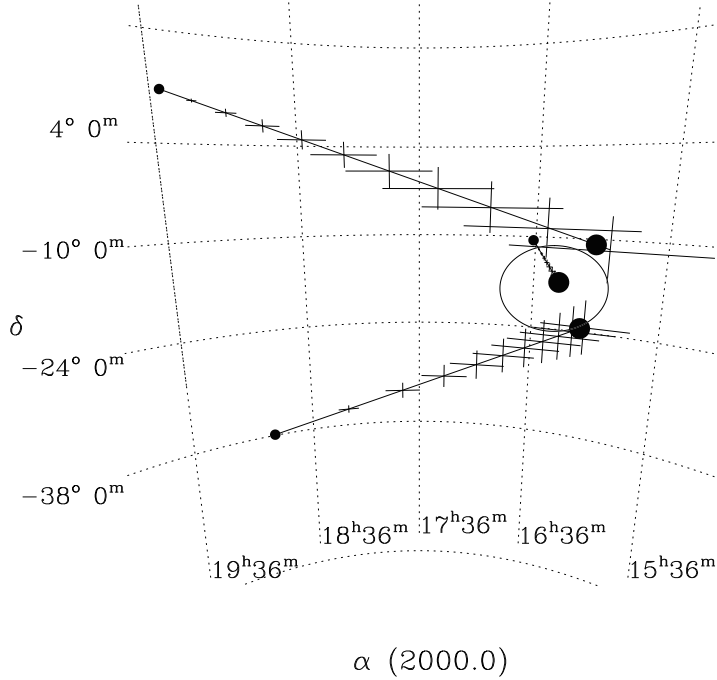,width=6.0in}}
\caption{Apparent positions of the
neutron star RX~J185635-3754 (moving ESE),
the pulsar B1929+10 (moving ENE), $\zeta$~Oph
(declination $-$10$^\circ$, moving NNE), and the
Upper Sco association (declination $-$24$^\circ$, moving SSW)
projected backwards in time. 
I use the assumed radial velocities of $-$60 and +160 km~s$^{-1}$
for RX~J185635-3754 and
PSR~B1929+10 which yield the closest approaches to $\zeta$~Oph.
The present positions are plotted as small solid dots; positions of
B1929+10, $\zeta$~Oph, and RX~J185635-3754 at times of
closest approach are the large filled circles. The large open circle is
the position and approximate size of the Upper Sco association at the time of
closest approach. The uncertainties on the positions are estimated using the
uncertainties in the distances, proper motions, and radial velocities, and
are plotted for the three stars at 10 points spaced linearly in time between
0.1 and 1.0 million years in the past.
}\label{fspm}
\end{figure}

\placefigure{fspm}

de Zeeuw et al.\/ (1999) catalog members of the Upper Sco and Upper Cen-Lup
associations, along with mean distances and radial velocities of the
associations. I went back to the listings for the individual
members in the Hipparcos Catalog (ESA 1997), and confirmed the mean proper
motions. de Zeeuw et al.\/ quote diameters of $\sim$14 and 27 degrees
for the two associations. Although the associations are almost certainly 
expanding and dispersing, I assume a constant radial extent in the analysis
below.

I take the parallax and proper motion of $\zeta$~Oph (HD~149757)
from the Hipparcos
Catalog (ESA 1997). I adopt the radial velocity of $-$9~km~s$^{-1}$ quoted in
the Hipparcos Input Catalog (Turon et al.\/ 1993). Evans (1979) quotes a radial
velocity of $-$15~km~s$^{-1}$. van Rensbergen et al.\/ (1996)
quote a peculiar velocity of
18~km~s$^{-1}$ with respect to the Sco-Cen OB association, apparently assuming
a single radial velocity for the association.
In light of the range in the radial velocities,
and the difficulty of establishing the radial velocity of a hot
star with such rapid rotation, I adopt an uncertainty of $\pm$5~km~s$^{-1}$.

I examine the space motions by converting all motions to heliocentric
Cartesian coordinates. Uncertainties in the
measurements are propagated through. 
I assume an uncertainty of 5~km~s$^{-1}$ in the radial velocity.
Uncertainties in the position due to
space motions increase linearly with time, but the uncertainty on the location
of the centers of the associations is dominated by their large radial extents.
The kinematic parameters I used are summarized in Table~\ref{tbl-input};
the space motions are shown in Figure~\ref{fspm}. 

\placetable{tbl-input}

\subsubsection{$\zeta$ Oph and the Sco-Cen OB Associations}

Blaauw (1991, 1993) noted that
$\zeta$~Oph appears to be moving directly away from the Upper Sco association.
Indeed, the space motion of $\zeta$~Oph takes it within 2.3$\pm$15~pc of the
center of the Upper Sco association 1.0~Myr ago. This
is well within the $\sim$7$^\circ$ (17~pc)
radial extent of the association at its
distance of 140~pc. Assuming a linear expansion with time, the cluster should
have been only about 20\% smaller at that time, so within the uncertainties
$\zeta$~Oph either passed through or originated near the center of the
Upper Sco association.

 However, van Rensbergen et al.\/ (1996) have used 
stellar evolutionary arguments to suggest that $\zeta$~Oph is older, and
originated in the Upper Cen-Lup association.
They note that the space motion of $\zeta$~Oph intersects
the location of the Upper Cen-Lup association 2-3 Myr in the past.
From age constraints, and by modelling the atmosphere of $\zeta$~Oph, van
Rensbergen et al.\/ argue that it cannot be a single star ejected from Upper Sco
by gravitational encounters with other massive stars. They could not find a
binary evolution scenario that produced a star like $\zeta$~Oph in less than
about 9 Myr, an age which greatly exceeds that of Upper Sco. They conclude it
likely that $\zeta$~Oph was a binary in the Upper Cen-Lup association, and that
it was ejected from that association when its companion star became a supernova
some 2-3 Myr ago.

The space motions of $\zeta$~Oph and the Upper Cen-Lup association cast
doubt on this scenario. The closest approach of the
two, 27~$\pm$~22~pc at 2.1~Myr, is only marginally consistent with
origination within the $\sim$30~pc (13$^\circ$) radius
of the association today (linear expansion suggests about a 25~pc radius
at 2.1~My). Note that the Hipparcos parallaxes of $\zeta$~Oph and the
two associations are virtually identical, while the radial velocities of
the associations differ by about 9~km~s$^{-1}$. The radial velocity of
$\zeta$~Oph is within about 5~km~s$^{-1}$ of that of Upper Sco, but is about
13~km~s$^{-1}$ less than that of the Upper Cen-Lup association.
If so,  two million years ago $\zeta$~Oph would have been 30~pc
further away from the Sun than was the Upper Cen-Lup association.

The $-$9~km~s$^{-1}$ radial velocity of $\zeta$~Oph minimizes
the closest approach to
the center of
the Upper~Sco association, while a radial velocity of +5~km~s$^{-1}$
is required to
bring the star with 5~pc of the center of Upper Cen-Lup. It is unlikely that the
radial velocity measurements will be this far off, so I conclude based on
geometry alone that it is not likely that $\zeta$~Oph originated in
Upper Cen-Lup.

\subsubsection{The Wanderings of the Neutron Star}

One cannot be so definitive about the space motions of the neutron star,
because its
radial velocity is unknown and not easily measurable. Rather, I
look for plausibility that the neutron star
originated in one of these associations,
or as companion to $\zeta$~Oph. For the measured proper motion and parallax,
the neutron star comes within 18$\pm$17~pc of $\zeta$~Oph at 1.0~Myr
and within 16$\pm$12~pc of the center of the
Upper~Sco association (radius about 7~pc) at 0.9~Myr, for a radial velocity of
$\sim-$55 to --60~km~s$^{-1}$. 
The closest approach of the neutron star to the Upper Cen-Lup association is
33$\pm$12~pc at 0.9~Myr, for a radial velocity of
$\sim-$45~km~s$^{-1}$.
These distances are summarized in Table~\ref{tbl-4}.

\placetable{tbl-4}

At 90\% confidence, for a radial velocity of $-$55 to $-$60 km~s$^{-1}$,
the positions of the neutron star, $\zeta$~Oph, and the center of the Upper~Sco
association all come together between 0.9 and 1.0~Myr ago. 
For no plausible combination of radial velocity and current distance do
the NS, $\zeta$~Oph, and the center of the Upper Cen-Lup association 
ever coincide within 90\% confidence.

It is unlikely that the spatial coincidence of $\zeta$~Oph, the Upper Sco
association, and the neutron star some 1.0~Myr in the past is a mere
coincidence.
Given a neutron star at the position of RX~J185635-3754, the likelihood that
a random motion will cause it to coincide {\it in projection only} with
the Upper Sco association ($\sim$7$^\circ$ radius) is about 0.007.
Therefore I identify the neutron star as the possible remnant of the
original primary of the $\zeta$~Oph system. 
That the temperature of the neutron star is close to that expected
from the standard neutron star cooling curve at 1~Myr may be further
circumstantial evidence supporting this identification.

\subsubsection{PSR B1929+10}

RX~J185635-3754 is not the only neutron star moving away from the general
direction of the Sco-Cen complex. The nearby pulsar PSR B1929+10
(PSR J1932+1059) is another candidate for ejection from Upper Sco
(Hoogerwerf, de Bruijne, \& de Zeeuw 2000). Downes \& Reichley (1983)
reported the proper motion. There are 3 available
distances (summarized by Pavlov et al.\/ 1996a); I adopt 
Backer \& Sramek's (1981) lower limit of 250~pc 
\footnote{The 170~pc distance suggested by the dispersion measure 
may be suspect because the galactic model does not account for small scale 
irregularities in the electron distribution (Taylor \& Cordes 1993).}.
The spindown age is 3~Myr. Applying the same
techniques used above, pulsar approaches to within 
19$\pm$30~pc of $\zeta$~Oph 1.0~Myr in the past, and is 22$\pm$27~pc from the
center of the Upper Sco association at about the same time, if the pulsar
is now at a  distance of 250$\pm$25~pc and has a
radial velocity of 160~km~s$^{-1}$. 
Clearly this pulsar is also a possible candidate for a
binary companion to $\zeta$~Oph, since the same arguments as put forth above
for the chance coincidence apply. Both cannot be the companion.
Perhaps there was more than one 
supernova in a binary system in Upper~Sco about 1 million years ago. 

\subsection{Kick Velocity of the Neutron Star}
If RX~J185635-3754 is indeed the former companion of $\zeta$~Oph, then
their space motions provide
sufficient information to estimate the kick velocity of the neutron star.
They are separating at an angle of 69$^\circ$.
In the frame of the Upper~Sco association, the space velocities of $\zeta$~Oph
and the neutron star are 34~km~s$^{-1}$ and 104~km~s$^{-1}$, respectively. 
I assume that the velocity of $\zeta$~Oph was its orbital velocity at the
time the supernova unbound the system.

Evolutionary models (e.g., van Rensbergen et al.\/ 1996) suggest that
$\zeta$~Oph has a mass of 20-25~M$_\odot$. If the immediate progenitor of
the supernova was a 4-5~M$_\odot$ helium star, then the mass ratio would have
been about 4 or 5 to 1, with $\zeta$~Oph the more massive star.
Ignoring the momentum of the 3-4 M$_\odot$ of material lost in the
explosion, the remnant should have a space velocity directed anti-parallel
to that of $\zeta$~Oph, with a speed higher by the mass ratio. To convert
this into the observed space velocity of the neutron star
requires a kick velocity
of about 200 km~s$^{-1}$ directed 25$^\circ$ from the direction of motion of
$\zeta$~Oph. A lower initial mass ratio requires a lower kick velocity; if the
star that exploded was twice the mass of $\zeta$~Oph, the kick velocity
is 90~km~s$^{-1}$ directed 60$^\circ$ from the direction of motion of
$\zeta$~Oph.

\section{Conclusions}
RX~J185635-3754 is confirmed to be an isolated neutron star, at a distance
of 61\ud{9}{8}~pc, with a heliocentric space velocity
of 108\ud{16}{14}~km~s$^{-1}$.
It appears to have left the
Upper Sco OB association between 0.9 and 1.0 million years ago.
It may have been the
binary companion of the runaway O star $\zeta$~Oph, which left Upper Sco at
the same time. If so, the neutron star suffered a kick velocity of about
200~km~s$^{-1}$ amplitude at birth. 

Reexamination of the space motions of $\zeta$~Oph and the Sco-Cen OB
association casts some doubt on van Rensenbergen et al.'s conclusion
that $\zeta$~Oph originated in the Upper-Cen-Lup association. It is
not clear how to reconcile their binary evolutionary scenarios with these
geometric constaints.

The existence of an old, isolated neutron star of known age permits one to
place another point on the cooling curve, a point not contaminated by possible
non-thermal emission. The exact temperature depends on the choice of
atmospheric model, but in any event the luminosity lies near the FP
(Friedman \& Pandharipande 1981) cooling curve at an age near 1~Myr.

The inferred radius of the neutron star depends on the angular diameter, which
is model-dependent. The smallest angular diameter for a given temperature
is given by a black body. For a temperature kT=49~eV, the lower bound on
the radius R$_\infty$ is 6.0\ud{1.2}{0.6}~km, and preliminary
atmospheric models yield R$_\infty$~=~11.2$\pm$3.4~km.

RX~J185635-3754 will make its closest approach to the Earth in about 280,000
years, at a distance of 52$\pm$9~pc, in the constellation Grus.

\acknowledgements
I thank J. Lattimer, M. Prakash,
R. Neuh\"auser, R. Wijers, R. Bandiera, and an anonymous referee
for useful comments and suggestions.
This research has been supported NASA through grants GO~064290195A and
GO~074080196A from the Space Telescope Science Institute, and by
LTSA grant NAG57978, to SUNY Stony Brook.

\clearpage

\clearpage
\begin{deluxetable}{rlrrr}
\tablewidth{0pt}
\tablecaption{WFPC2 Observation Log\label{tbl-1}}
\tablehead{
 \colhead{Program} & \colhead{Date} & \colhead{Root} & \colhead{Roll} &
                     \colhead{time}\\
  & \colhead{(UT)} & & \colhead{(deg)} & \colhead{(sec)}  }
\startdata
6149 & 1996 Oct  6 & U3IM01 & 129.5 & 4800 \\
7408 & 1999 Mar 30 & U51G01 & $-$51.6 & 7200 \\
7408 & 1999 Sep 16 & U51G02 & 124.2& 5191 
\enddata
\end{deluxetable}

\clearpage
\begin{deluxetable}{rrrrl|rrrrl}
\footnotesize
\tablecaption{Objects Identified in WFPC2 Images\tablenotemark{a}\label{tbl-2}}
\tablehead{
\colhead{ID} & \colhead{RA (s)} & \colhead{DEC (\arcsec)} & 
               \colhead{M$_{F606W}$} & \colhead{Note\tablenotemark{b}} &
\colhead{ID} & \colhead{RA (s)} & \colhead{DEC (\arcsec)} & 
               \colhead{M$_{F606W}$} & \colhead{Note\tablenotemark{b}} \\
 & \colhead{+18$^{\rm h}$ 56$^{\rm s}$} & 
   \colhead{$-$(37$^\circ$ 54$^{\arcmin}$)} &
 & & & \colhead{+18$^{\rm h}$ 56$^{\rm s}$} & 
       \colhead{$-$(37$^\circ$ 54$^{\arcmin}$)} 
}
\startdata
100 & 34.2647 & 29.977 &  25.1 & & 122 & 35.9440 & 35.188 &  26.3 \\ 
101 & 34.5549 & 23.679 &  24.2 & & 123 & 36.3414 & 37.038 &  26.4 \\ 
102 & 34.5906 & 22.338 &  24.4 & & 124 & 35.5635 & 40.923 &  26.8 \\ 
103 & 34.8373 & 16.947 &  23.8 & & 125 & 35.5023 & 32.800 &  26.6 \\ 
104 & 34.9456 & 16.308 &  25.4 & E & 126 & 35.8937 & 28.822 &  26.8 \\ 
105 & 35.4770 & 21.313 &  25.9 & & 127 & 35.3542 & 49.328 &  25.9 \\ 
106 & 35.5512 & 25.310 &  22.3 & & 128 & 35.2637 & 47.997 &  24.9 \\ 
107 & 34.4844 & 43.087 &  23.5 & & 129 & 35.0718 & 44.297 &  26.2 \\ 
108 & 35.8185 & 47.302 &  25.2 & E & 130 & 35.1307 & 14.480 &  26.4 \\ 
109 & 35.5538 & 42.777 &  25.7 & E & 131 & 35.1498 & 17.022 &  26.4 \\ 
110 & 36.0382 & 46.799 &  25.0 & & 132 & 34.4582 & 25.389 &  26.3 \\ 
111 & 37.0722 & 29.560 &  24.5 & & 133 & 35.3043 & 29.448 &  20.2 \\
112 & 36.3679 & 24.570 &  23.6 & & 19 & 34.2052 & 40.171 &  21.9 & 1 \\ 
113 & 36.1096 & 33.936 &  23.8 & & 20 & 35.0369 & 39.048 &  24.0 & 1; E \\ 
114 & 35.9229 & 31.175 &  24.2 & & 21 & 34.7937 & 37.178 &  20.9 & 1\\ 
115 & 35.7030 & 37.480 &  26.5 & & 23 & 35.1855 & 36.142 &  23.7 & 1\\ 
116 & 33.6869 & 36.232 &  24.0 & & 24 & 35.9404 & 33.506 &  20.3 & 1\\ 
117 & 34.5558 & 23.072 &  25.7 & & 26 & 36.8333 & 29.630 &  21.4 & 1\\ 
118 & 34.9533 & 25.884 &  24.6 & & 28 & 34.8730 & 29.639 &  21.5 & 1\\ 
119 & 35.6693 & 24.284 &  25.7 & &  I & 35.2597 & 14.740 &  18.9 & 2\\ 
121 & 36.0028 & 35.401 &  26.0 & &  J & 34.7934 & 25.762 &  20.9 & 2\\ 
  X & 35.5003 & 36.823 &  25.9 & 3 \\ 
\enddata
\tablenotetext{a}{These positions are measured on the U3IM01 image.
                  The coordinates are equinox J2000, epoch 1996.75.}
\tablenotetext{b}{(1): ID from Campana et al.\/ 1997;
                  (2): ID from Walter et al.\/ 1996.;
                  (3): the target, RX J185635-3754;
                  E: extended source, probably a galaxy.}
\normalsize
\end{deluxetable}

\clearpage
\begin{deluxetable}{rrrrr}
\tablewidth{0pt}
\tablecaption{The Motions of RX~J185635-3754\label{tbl-3}}
\tablehead{
 \colhead{Quantity} & & \colhead{68\% confidence} & \colhead{95\% confidence}}

\startdata
Proper Motion ($\alpha$) & 326.7 & $\pm$ 0.8 & $\pm$ 1.5 & mas~yr$^{-1}$ \\
Proper Motion ($\delta$) & $-$59.1 & $\pm$ 0.7 & $\pm$ 2.2 & mas~yr$^{-1}$ \\
Proper Motion            & 332.0 & \ud{1.1}{1.3} & \ud{1.8}{2.0} & mas~yr$^{-1}$ \\
Position Angle           & 100.3 & $\pm$ 0.1 & $\pm$ 0.3 & $^\circ$ \\
Parallax                 &  16.5 & \ud{2.4}{2.2} & \ud{3.7}{3.5} & mas \\
\enddata
\end{deluxetable}

%

\clearpage
\begin{deluxetable}{rrrrl}
\tablewidth{0pt}
\tablecaption{Space Motions of $\zeta$ Oph and the Sco Cen
              Associations\tablenotemark{a}\label{tbl-input}}
\tablehead{
 \colhead{Quantity} & \colhead{$\zeta$ Oph} & \colhead{Upper Sco} &
  \colhead{Upper Cen-Lup} & }
\startdata
Distance                & 140 \ud{16}{12} &145 $\pm$ 3&142 $\pm$ 2&pc\\
Proper Motion ($\alpha$) &
     13.07 $\pm$ 0.85& $-$10.89 $\pm$ 2.3&$-$21.06 $\pm$ 5.1& mas~yr$^{-1}$ \\
Proper Motion ($\delta$) &
     25.44 $\pm$ 0.72& $-$23.39 $\pm$ 3.6&$-$23.35 $\pm$ 3.8& mas~yr$^{-1}$ \\
radial velocity          &
        $-$9 $\pm$ 5 & $-$4.6 $\pm$1.7 & 4.9 $\pm$1.7 & km~s$^{-1}$\\
\enddata
\tablenotetext{a}{Sources are referenced in the text.}
\end{deluxetable}

\clearpage
\begin{deluxetable}{lcrrr}
\tablewidth{0pt}
\tablecaption{Close Approaches\label{tbl-4}}
\tablehead{
\colhead{Encounter} & \colhead{Distance} & \colhead{Time} &
                      \colhead{RV$_{NS}$\tablenotemark{a}} &
                      \colhead{D$_{NS}$\tablenotemark{b}}\\
 & \colhead{(pc)} & \colhead{(Myr)} & \colhead{(km~s$^{-1}$)} &
   \colhead{(pc)}  }
\tablecolumns{5}
\startdata
$\zeta$~Oph - Upper Sco & 2$\pm$15 & 1.0 & \nodata & \nodata \\
$\zeta$~Oph - Upper Cen-Lup & 27$\pm$22 & 2.1 & \nodata & \nodata \\
\cutinhead{For nominal neutron star distance of 61 pc}
NS -$\zeta$~Oph      & 18$\pm$17 & 1.0   &  $-$55 & \nodata   \\
NS - Upper Sco       & 16$\pm$12 & 0.9   &  $-$60 & \nodata   \\
NS - Upper Cen-Lup   & 33$\pm$12 & 0.9   &  $-$45 & \nodata   \\
\cutinhead{For unconstrained neutron star distance}
NS-$\zeta$~Oph      &  1$\pm$22 & 2.1   &  $-$45 & 31  \\
NS- Upper Sco       &  7$\pm$10 & 0.4   &  +30 & 130  \\
NS- Upper Cen-Lup   &  1$\pm$22 & 2.4   &  $-$35 & 22  \\
\enddata
\tablenotetext{a}{ The neutron star (NS) radial velocity which minimizes the
                  distance of closest approach.}
\tablenotetext{b}{ The present distance of the neutron star which minimizes the
                  distance of closest approach.}
\end{deluxetable}

\end{document}